\documentclass[aps,prb,superscriptaddress,amsmath,amssymb,floatfix,showpacs]{revtex4}
\usepackage{graphicx}
\usepackage{subfigure}
\usepackage{ulem}
\usepackage{natbib}

\begin{document}

\title{Dynamics of a two-level system strongly coupled to a high-frequency quantum oscillator}

\author{E. K. Irish}
\email{eirish@pas.rochester.edu}
\affiliation{Department of Physics and Astronomy, University of Rochester, Rochester, New York 14627}
\affiliation{Laboratory for Physical Sciences\\
8050 Greenmead Drive, College Park, Maryland 20740}

\author{J. Gea-Banacloche}
\affiliation{Department of Physics, University of Arkansas, Fayetteville, Arkansas 72701}

\author{I. Martin}
\affiliation{Theoretical Division, Los Alamos National Laboratory, Los Alamos, New Mexico 87545}

\author{K. C. Schwab}
\affiliation{Laboratory for Physical Sciences\\
8050 Greenmead Drive, College Park, Maryland 20740}

\date{\today}

\begin{abstract}
Recent experiments on quantum behavior in microfabricated solid-state systems suggest tantalizing connections to quantum optics. Several of these experiments address the prototypical problem of cavity quantum electrodynamics: a two-level system coupled to a quantum harmonic oscillator. Such devices may allow the exploration of parameter regimes outside the near-resonance and weak-coupling assumptions of the ubiquitous rotating-wave approximation (RWA), necessitating other theoretical approaches. One such approach is an adiabatic approximation in the limit that the oscillator frequency is much larger than the characteristic frequency of the two-level system. A derivation of the approximation is presented, together with a discussion of its applicability in a system consisting of a Cooper-pair box coupled to a nanomechanical resonator. Within this approximation the time evolution of the two-level-system occupation probability is calculated using both thermal- and coherent-state initial conditions for the oscillator, focusing particularly on collapse and revival phenomena. For thermal-state initial conditions parameter regimes are found in which collapse and revival regions may be clearly distinguished, unlike the erratic evolution of the thermal-state RWA model. Coherent-state initial conditions lead to complex behavior, which exhibits sensitive dependence on the coupling strength and the initial amplitude of the oscillator state. One feature of the regime considered here is that closed-form evaluation of the time evolution may be carried out in the weak-coupling limit, which provides insight into the differences between the thermal- and coherent-state models. Finally, potential experimental observations in solid-state systems, particularly the Cooper-pair box--nanomechanical resonator system, are discussed and found to be promising.
\end{abstract}

\maketitle

\section{Introduction}\label{Intro}

One of the simplest fully quantum-mechanical systems consists of a harmonic oscillator coupled to a two-level (spin-like) system. Introduced in 1963 by Jaynes and Cummings, \cite{Jaynes:1963} this model has yet to be completely solved despite its apparent simplicity. The approach taken by the original authors, known as the rotating-wave approximation (RWA), relies upon the assumptions of near-resonance and weak coupling between the two systems. The RWA is widely used because it is readily solvable and describes quite accurately the standard physical realization of such a system: an atom coupled to a field mode of an electromagnetic cavity. In this experimental situation, the coupling strength between the atom and the field is largely determined by the intrinsic dipole moment of the atom; for all experiments to date, the coupling strength $\lambda$ is very small compared to the atomic transition frequency $\omega$ ($\lambda/\omega \sim 10^{-7}$--$10^{-6})$. \cite{Raimond:2001,Hood:2000} The near-resonance condition is necessary to ensure validity of the two-level description of the atom. Thus the RWA is a natural, and excellent, approximation in such a system.

Quantum-limited solid state devices offer an alternative to the traditional atom--cavity implementation of the spin--oscillator system. Recent experiments have shown clear spectroscopic evidence that a Cooper-pair box (CPB), or Josephson charge qubit, coupled to a superconducting transmission line behaves much like an atom in a cavity. The dipole coupling between the two systems is $\lambda/\omega \approx 10^{-3}$, $3-4$ orders of magnitude larger than that achieved in atomic systems. \cite{Wallraff:2004, Wallraff:2005} Capacitive or inductive couplings offer the possibility of still larger coupling strengths than those possible with dipole coupling, even at large detunings between the fundamental frequencies of the oscillator and the two-level system. Some results from a flux-based, inductively-coupled system give preliminary evidence for coupled quantum behavior and entanglement between the two-level system and the oscillator. \cite{Chiorescu:2004} Another intriguing possibility involves capacitively coupling a CPB \cite{Armour:2002a,Armour:2002b,Irish:2003} or a Josephson phase qubit \cite{Sornborger:2004} to a nanomechanical resonator (NR).  All of these systems are capable of accessing coupling strengths and detunings outside the regime in which the RWA is valid, requiring different theoretical approaches to the problem.  For example, Ref.~\onlinecite{Sornborger:2004} details a perturbative treatment which is valid for moderately strong coupling ($\lambda/\omega \lesssim 0.3$) at zero detuning. 

In this paper we discuss an approximation which is able to treat strong coupling and large detuning. It is valid when the splitting frequency of the two-level system is much smaller than the frequency of the oscillator and holds well even for coupling strengths up to or larger than the oscillator frequency. The approximation is used to examine the time evolution of the two-level system when the harmonic oscillator begins in a thermal state or a coherent state. Several effects of the coupling to the oscillator are distinguished, including enhanced apparent decoherence rates, frequency modification, and collapse and revivals of Rabi oscillations. We focus particularly on collapses and revivals in this model.

In the next section we introduce the form of the Hamiltonian to be considered and derive the adiabatic approximation. In Sec.\ \ref{Dynamical} we evaluate the approximate time evolution of the two-level system coupled to, respectively, a Fock state, a thermal state, and a coherent state of the oscillator, and classify the behavior in several parameter regimes. Section \ref{Weak coupling} contains a further approximation which allows evaluation in closed form of the infinite sums encountered in the thermal- and coherent-state models. The prospects for experimental observation of our predictions are analyzed in Sec.\ \ref{Experimental}, and Sec.\ \ref{Conclusion} concludes the paper. The Appendix contains a detailed derivation of the effective Hamiltonian for a CPB-NR system.

\section{Adiabatic approximation in the displaced oscillator basis}\label{Adiabatic}

The Hamiltonian which forms the basis for the calculations and discussion in this paper is
\begin{equation}\label{Hamiltonian}
H = \tfrac{1}{2}\hbar \Omega \hat{\sigma}_x + \hbar \lambda \hat{\sigma}_z (\hat{a}^{\dag} + \hat{a}) + \hbar \omega_0 \hat{a}^{\dag} \hat{a} .
\end{equation}
The Hamiltonian typically considered in cavity quantum electrodynamics \cite{Jaynes:1963,Shore:1993} (CQED) differs from Eq.\ \eqref{Hamiltonian} only by a rotation about $\hat{\sigma}_y$. Our notation is chosen based on the particular solid-state system which we have in mind, consisting of a Cooper-pair box coupled to a nanomechanical resonator. When the CPB is biased away from its degeneracy point and driven by a classical field resonant with the qubit transition frequency, an effective Hamiltonian for the coupled system may be found which has exactly the form of Eq.\ \eqref{Hamiltonian} (see Sec.\ \ref{Experimental} and the Appendix). \footnote{The same Hamiltonian may be obtained by biasing the CPB at its degeneracy point; however, in this case, typical experimental parameters result in $\Omega \gg \omega_0$, while the approximation discussed in this paper works in the opposite limit. See, for instance, Ref.\ \onlinecite{Irish:2003}.} In this case the two-level system undergoes Rabi oscillations due to the (classical) external driving field which are then altered by the coupling to the quantum oscillator, unlike the usual picture in which the two-level system is driven by the quantum oscillator itself. Although the Hamiltonian and therefore the results we obtain are not restricted to the CPB-NR system, for concreteness we will use the language of that system throughout most of the paper.

As no analytic solution to Eq.\ \eqref{Hamiltonian} is yet known, some approximation is required. The most common approach is to assume that the two-level system and the oscillator are close to resonance, ($\Omega - \omega_0) \ll \Omega, \omega_0$, and that the coupling between them is weak, $\lambda \ll \Omega, \omega_0$. Then terms which do not conserve energy may be discarded (the RWA), resulting in what is usually termed the Jaynes-Cummings model (JCM). \cite{Jaynes:1963,Shore:1993} As mentioned in the Introduction, this is an appropriate approximation for atom-cavity experiments; however, unlike atom-cavity systems, the solid-state system we are considering has the potential for strong coupling at large detunings. In this paper we will treat the case in which the two-level-system splitting frequency is much smaller than the oscillator frequency, $\Omega \ll \omega_0$, and the coupling strength is allowed to be large, on the order of or greater than the oscillator frequency. The RWA is not appropriate in this limit.

An excellent way to treat this regime is via a type of adiabatic approximation. This approximation has been derived previously by several authors using different methods. Graham and H{\"o}hnerbach refer to this regime as the ``quasi-degenerate limit'' and give the same lowest-order expressions as we derive. \cite{Graham:1984a,Graham:1984b,Graham:1984c} Schweber utilized the Bargmann Hilbert-space representation \cite{Schweber:1967} and Crisp solved recurrence relations; \cite{Crisp:1992} both of these authors found higher-order corrections beyond what we present. We take yet a different approach. First, by neglecting the self-energy of the two-level system, we derive the basis in which the rest of the calculation will be performed, called the ``displaced oscillator'' \cite{Crisp:1992} basis. In this basis the Hamiltonian may be truncated to a block diagonal form and the blocks solved individually. Essentially, the two-level-system self-energy is introduced only as needed to lift the degeneracy within individual subspaces in the displaced oscillator basis. 

\begin{figure}
\includegraphics[scale=1]{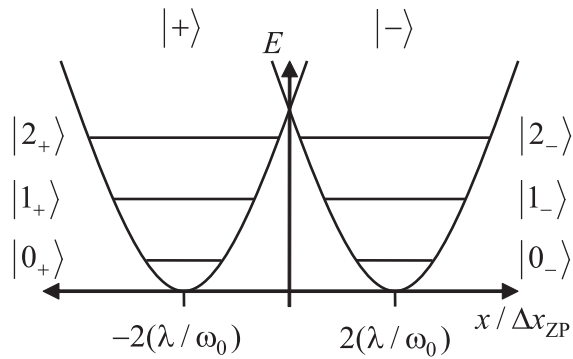}
\caption{\label{fig:displbasis1} Graphical representation of the displaced oscillator basis. The equilibrium position of the harmonic oscillator well is shifted by an amount proportional to the coupling constant $\lambda$, with the direction of the shift determined by the state of the two-level system. Each well retains its harmonic character, with the usual eigenstates. Eigenstates with the same value of $N$ are degenerate in energy.}
\end{figure}

To begin with we consider eigenstates of the form $\lvert i, \phi_i \rangle = \lvert i \rangle \otimes \lvert \phi_i \rangle$ where $\lvert i = +,-\rangle$ denotes the eigenstates of $\hat{\sigma}_z$ and $\lvert \phi_{i} \rangle$ are the corresponding oscillator eigenstates, found from the last two terms of Eq.\ \eqref{Hamiltonian} with $\hat{\sigma}_z$ set to its eigenvalue of $\pm 1$ as appropriate:
\begin{equation}
[\pm \hbar\lambda (\hat{a}^{\dag} + \hat{a}) + \hbar\omega_0 \hat{a}^{\dag} \hat{a}] \lvert \phi_{\pm} \rangle = E \lvert \phi_{\pm} \rangle .
\end{equation}
Completing the square gives
\begin{equation} 
\left[\left(\hat{a}^{\dag} \pm \frac{\lambda}{\omega_0}\right)\left(\hat{a} \pm \frac{\lambda}{\omega_0}\right)\right] \lvert \phi_{\pm} \rangle= \left(\frac{E}{\hbar\omega_0} + \frac{\lambda^2}{\omega_0^2}\right) \lvert \phi_{\pm} \rangle .
\end{equation}
Taking $\lambda / \omega_0$ to be real, the operator on the left-hand side may be rewritten as
\begin{equation}\label{displnumop}
\left[\left(\hat{a}^{\dag} \pm \frac{\lambda}{\omega_0}\right)\left(\hat{a} \pm \frac{\lambda}{\omega_0}\right)\right] = \hat{D}(\mp \lambda/\omega_0) \hat{a}^{\dag} \hat{a} \hat{D}^{\dag}(\mp \lambda/\omega_0) ,
\end{equation}
where $\hat{D}(\nu)=\exp[\nu(\hat{a}^{\dag} - \hat{a})]$ is a displacement operator. \cite{Mandel:optical} The operator in Eq.\ \ref{displnumop} may be interpreted as the number operator for a harmonic oscillator of frequency $\omega_0$, effective mass $m$, and equilibrium position $\mp (2 \lambda/\omega_0) \Delta x_{\text{ZP}}$ where $\Delta x_{\text{ZP}} = \sqrt{\hbar/(2 m \omega_0)}$. The eigenstates of this operator are displaced Fock (number) states,
\begin{equation}\label{displstates}
\lvert \phi_{\pm} \rangle = e^{\mp (\lambda / \omega_0)(\hat{a}^{\dag} - \hat{a})} \lvert N \rangle \equiv \lvert N_{\pm} \rangle \qquad N=0,1,2,\dotsc 
\end{equation}
with energies given by
\begin{equation}\label{displener}
E_N = \hbar\omega_0\left(N - \frac{\lambda^2}{\omega_0^2}\right) .
\end{equation}
These approximate eigenstates and energies constitute the displaced oscillator basis, which will be used throughout subsequent calculations. An illustration of the displaced-well potentials and the corresponding energy levels is given in Fig.\ \ref{fig:displbasis1}.

This basis has a simple interpretation in the CPB-NR system. Due to the capacitive coupling, one charge state of the CPB will attract the NR and shift its equilibrium position closer to the CPB, while the oppposite charge state will repel the NR. As long as the stretching of the beam due to the attraction is small, the NR will remain approximately linear and may still be described as a harmonic oscillator. However, the position of the harmonic potential well will have shifted. This is precisely the situation described mathematically by the displaced oscillator basis, which allows the use of the ordinary harmonic oscillator formalism within each displaced potential well. The correspondence of physical position displacement to the formal displacement operator makes this a natural basis to work in.

However, there are some peculiarities associated with using the displaced oscillator states as a basis. Although the states $\lvert +, N_{+} \rangle$ ($\lvert -, N_{-} \rangle$) form an orthonormal basis with $\langle M_+ \vert N_+ \rangle = \delta_{MN}$ ($\langle M_- \vert N_- \rangle = \delta_{MN})$, the states $\lvert N_{+} \rangle$ and $\lvert N_{-} \rangle$ are not mutually orthogonal. This may be understood easily in position space, where the displacement operator corresponds to a displacement in $x$. Consider, for example, the harmonic oscillator ground state $\lvert 0 \rangle$, which has a Gaussian form: this state displaced by a finite amount is never completely orthogonal to the same state displaced by the opposite amount, although the displaced states may become very nearly orthogonal if the displacement is large enough. The overlap between Fock states displaced in different directions is given by
\begin{equation}\label{overlap}
\langle M_- \vert N_+ \rangle = \begin{cases}
e^{-2 \lambda^2 / \omega_0^2} (-2 \lambda / \omega_0 )^{M-N} \sqrt{N! / M!}~L_{N}^{M-N}[(2 \lambda / \omega_0)^2]&  M \ge N \\
e^{-2 \lambda^2 / \omega_0^2} (2 \lambda / \omega_0)^{N-M} \sqrt{M! / N!}~L_{M}^{N-M}[(2 \lambda / \omega_0)^2]&  M < N ,
\end{cases}
\end{equation}
where $L_i^j$ is an associated Laguerre polynomial. Note that $\langle M_- \vert N_+ \rangle = (-1)^{N-M} \langle N_- \vert M_+ \rangle$. Together with the fact that $\langle M_- \vert N_+ \rangle$ is real, this implies that $\langle M_+ \vert N_- \rangle = (-1)^{M-N} \langle M_- \vert N_+ \rangle$, which is a useful identity. The lack of orthogonality between different displacements leads to the unusual results in the two-level-system dynamics which will be found later.

The next step is to reintroduce the two-level-system self-energy, $\Omega$, which is assumed to be small. Reformulating the problem in terms of a matrix written in the displaced oscillator basis allows the approximation to be carried out in a natural way. Using the overlap functions calculated above, the matrix representing the Hamiltonian~\eqref{Hamiltonian} may be written down:
\begin{equation}\label{matrix1}
H = 
\begin{pmatrix}
E_0 & \tfrac{\Omega}{2}\langle 0_- \vert 0_+ \rangle & 0 & \tfrac{\Omega}{2}\langle 1_- \vert 0_+ \rangle & 0 & \tfrac{\Omega}{2}\langle 2_- \vert 0_+ \rangle & \dots \\
\tfrac{\Omega}{2} \langle 0_- \vert 0_+ \rangle & E_0 & -\tfrac{\Omega}{2}\langle 1_- \vert 0_+ \rangle & 0 & \tfrac{\Omega}{2} \langle 2_- \vert 0_+ \rangle & 0 & \dots \\
0 & -\tfrac{\Omega}{2} \langle 1_- \vert 0_+ \rangle & E_1 & \tfrac{\Omega}{2} \langle 1_- \vert 1_+ \rangle & 0 & \tfrac{\Omega}{2} \langle 2_- \vert 1_+ \rangle & \dots \\
\tfrac{\Omega}{2} \langle 1_- \vert 0_+ \rangle & 0 & \tfrac{\Omega}{2} \langle 1_- \vert 1_+ \rangle & E_1 & -\tfrac{\Omega}{2} \langle 2_- \vert 1_+ \rangle & 0 & \dots \\ 
0 & \tfrac{\Omega}{2} \langle 2_- \vert 0_+ \rangle & 0 & -\tfrac{\Omega}{2} \langle 2_- \vert 1_+ \rangle & E_2 & \tfrac{\Omega}{2} \langle 2_- \vert 2_+ \rangle & \dots \\
\tfrac{\Omega}{2} \langle 2_- \vert 0_+ \rangle & 0 & \tfrac{\Omega}{2} \langle 2_- \vert 1_+ \rangle & 0 & \tfrac{\Omega}{2} \langle 2_- \vert 2_+ \rangle & E_2 & \dots \\
\vdots & \vdots & \vdots & \vdots & \vdots & \vdots & \ddots
\end{pmatrix} ,
\end{equation}
where the order of the columns and rows is $\lvert +,0_+ \rangle, \lvert -,0_- \rangle, \lvert +,1_+ \rangle, \lvert -,1_- \rangle, \ldots$. The approximation consists of truncating the matrix \eqref{matrix1} to the block diagonal form
\begin{equation}\label{matrixapprox}
H \approx 
\begin{pmatrix}
E_0 & \tfrac{\Omega}{2}\langle 0_- \vert 0_+ \rangle & 0 & 0 & 0 & 0 & \dots \\
\tfrac{\Omega}{2} \langle 0_- \vert 0_+ \rangle & E_0 & 0 & 0 & 0 & 0 & \dots \\
0 & 0 & E_1 & \tfrac{\Omega}{2} \langle 1_- \vert 1_+ \rangle & 0 & 0 & \dots \\
0 & 0 & \tfrac{\Omega}{2} \langle 1_- \vert 1_+ \rangle & E_1 & 0 & 0 & \dots \\ 
0 & 0 & 0 & 0 & E_2 & \tfrac{\Omega}{2} \langle 2_- \vert 2_+ \rangle & \dots \\
0 & 0 & 0 & 0 & \tfrac{\Omega}{2} \langle 2_- \vert 2_+ \rangle & E_2 & \dots \\
\vdots & \vdots & \vdots & \vdots & \vdots & \vdots & \ddots
\end{pmatrix} .
\end{equation}
In this approximation, the self-energy of the two-level system has been employed only where it is needed to lift the degeneracy of the displaced-oscillator basis states [Eqs.\ \eqref{displstates} and \eqref{displener}; also see Fig.\ \ref{fig:displbasis1}]. The natural interpretation of this form of the Hamiltonian is that the oscillator will be restricted to remain in the $N_{\pm}$ subspace if it has been initialized in $\lvert N_{+} \rangle$ or $\lvert N_{-} \rangle$. In other words, mixing occurs only between levels in opposite wells which have the same value of $N$ and thus the same energy. In this picture the condition $\Omega \ll \omega_0$ may be seen as a statement about statics: the spacing of the oscillator energy levels is very large compared to the spacing of the two-level system, so a transition in the two-level system can never excite the oscillator. An alternative argument, found in Sec.\ III(C) of Ref.\ \onlinecite{Leggett:1987}, relies on the separation of characteristic times in the two systems: given the assumption $\Omega \ll \omega_0$, the oscillator responds almost instantaneously to changes in $\langle \hat{\sigma}_z \rangle$, so that the oscillator dynamics is ``slaved'' to the two-level-system dynamics. 

\begin{figure}
\includegraphics[scale=1]{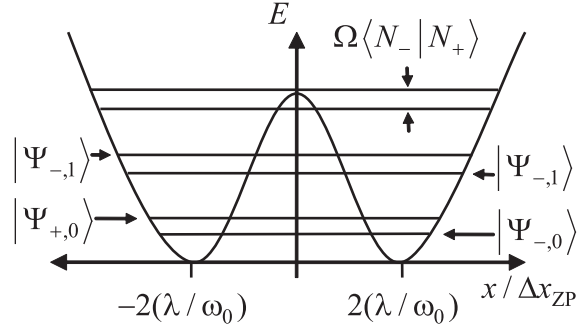}
\caption{\label{fig:displbasis2} Graphical representation of the adiabatic approximation. The two displaced oscillator wells, illustrated in Fig.\ \ref{fig:displbasis1}, are now allowed to interact. Levels with the same value of $N$ become mixed; the resultant energy splitting is proportional to the overlap of the wavefunctions.}
\end{figure}

Due to its simple block diagonal form, Eq.\ \eqref{matrixapprox} may be solved easily. The solutions in the $N$th block are given by
\begin{subequations}\label{eigen1}
\begin{gather}
\lvert \Psi_{\pm, N} \rangle = \tfrac{1}{\sqrt{2}}(\lvert +, N_+ \rangle \pm \lvert -, N_- \rangle) , \label{states1} \\
E_{\pm, N} = \pm \frac{1}{2} \Omega \langle N_- \vert N_+ \rangle + E_N . \label{energy1} 
\end{gather}
\end{subequations} 
These energies and eigenstates constitute the adiabatic approximation to lowest order in $\Omega/\omega_0$. A graphical representation is shown in Fig.\ \ref{fig:displbasis2}.

\begin{figure*}
\includegraphics[scale=1]{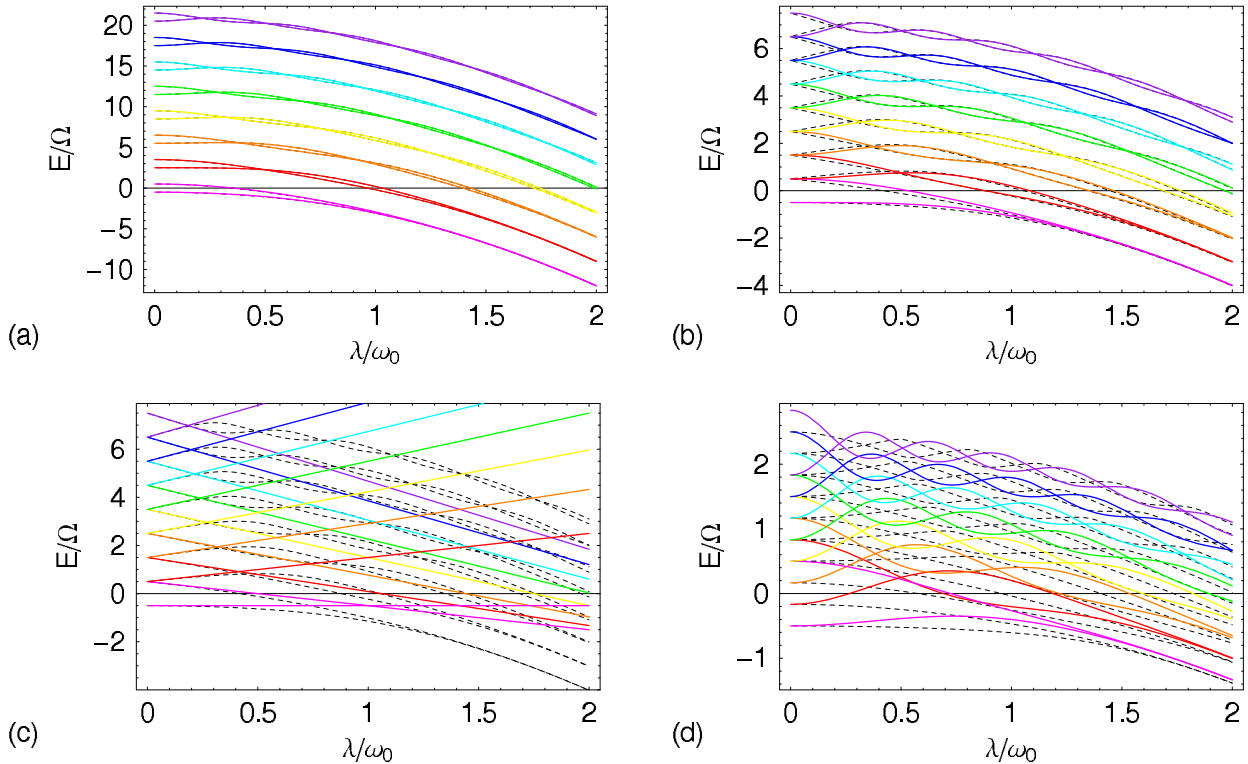}
\caption{\label{fig:energies}(Color online) Energy levels given by analytic approximation methods (solid lines) and by numerical solution of the full Hamiltonian (dashed lines). (a) Adiabatic approximation and numerical solution with $\Omega/\omega_0 = 1/3$. (b) Adiabatic approximation and numerical solution with $\Omega/\omega_0 = 1$. (c) Rotating wave approximation and numerical solution with $\Omega/\omega_0 = 1$. (d) Adiabatic approximation and numerical solution with $\Omega/\omega_0 = 3$.}
\end{figure*}

Figure~\ref{fig:energies} plots the lowest-lying energy levels given in Eq.\ \eqref{energy1} as a function of the coupling strength $\lambda/\omega_0$ for different values of the ratio $\Omega/\omega_0$. For comparison purposes, the results of a numerical diagonalization of the full Hamiltonian [Eq.\ \eqref{Hamiltonian}] are also shown. 

Consider first the case $\Omega/\omega_0=1/3$, shown in Fig.\ \ref{fig:energies}(a). This is within the regime in which the approximate solution \eqref{eigen1} should be valid. The ``ripple'' structure imposed by the Laguerre polynomials on the smooth variation of $E$ with $\lambda/\omega_0$ is immediately evident. This structure was noted in Ref.\ \onlinecite{Crisp:1992} and interpreted as an interference between states displaced in opposite directions. In the limit $\lambda/\omega_0 \to \infty$ the distance between the wells becomes infinite, and the overlap $\langle N_- \vert N_+ \rangle \to 0$. With no mixing between the wells the spectrum becomes that of two identical harmonic oscillators and the energy levels become pairwise degenerate. Agreement with the numerical solution is excellent.

Another noteworthy feature of the structure of the energy levels is the multiple crossings which appear between pairs of levels. These are true crossings, not narrow avoided crossings, allowed by conservation of the parity operator $\hat{P} = \exp[i \pi (\hat{a}^{\dag} \hat{a} + \tfrac{1}{2} + \tfrac{1}{2}\hat{\sigma}_x)]$ in Eq.\ \eqref{Hamiltonian}. \cite{Graham:1984a} The approximate eigenstates \eqref{states1} are eigenstates of $\hat{P}$; pairs of levels with different eigenvalues of $\hat{P}$ are allowed to cross. In other words, at the crossing points the quantum states in the two displaced wells destructively interfere with one another, destroying the tunneling process which mixes the states and provides the energy splitting.

Figure~\ref{fig:energies}(b) shows the resonance case, $\Omega/\omega_0=1$. Plots for this case also appear in Ref.\ \onlinecite{Graham:1984b}. Interestingly, the agreement between the approximate solution and the numerical solution is still quite good, especially at larger coupling strengths. Compare this to the RWA energy levels, \cite{Jaynes:1963} $E_{\pm,N}^{\text{RWA}} = N \hbar\omega_0 \pm \Omega/2 \pm \hbar \lambda \sqrt{N}$, shown in Fig.\ \ref{fig:energies}(c). The RWA gives the correct limiting behavior as $\lambda/\omega_0 \to 0$, but diverges from the numerical solution starting around the point where the paired levels first cross. This comparison illustrates the dependence of the RWA upon the assumption of weak coupling, even when the oscillator and two-level system are exactly resonant. 

Finally, Fig.\ \ref{fig:energies}(d) shows the case $\Omega/\omega_0=3$, which is out of the regime in which the adiabatic approximation is expected to hold. Plots of this case are given in Ref.\ \onlinecite{Graham:1984a}; however, the approximate and numerical solutions are shown in different figures, making it difficult to compare the two. The plot shown here demonstrates that the adiabatic approximation breaks down in this regime except in the broadest qualitative sense. When $\Omega/\omega_0>1$ spurious level crossings appear at small values of $\lambda/\omega_0$, and a substantial phase difference develops between the ripples in the numerical and approximate solutions. Thus it may be seen that, as expected, the adiabatic approximation is not a reasonable treatment for $\Omega/\omega_0 > 1$.

The displaced oscillator basis provides a physically intuitive picture for the derivation of an adiabatic approximation in the regime $\Omega/\omega_0 \le 1$. Comparing the adiabatic approximation with a numerical solution of the full Hamiltonian shows that the approximation works quite well in the regime for which it was derived. Although this approximation has been derived previously, no authors seem to have explored its consequences for experimental observables. Much of the remainder of this paper will be devoted to a study of those consequences.

\section{Dynamical behavior}\label{Dynamical}

For applications in real systems, the dynamical behavior of the two-level system is of particular interest. In this section we discuss the time dependence of the two-level-system observable $\hat \sigma_z$, which corresponds to charge in a Cooper-pair box. Three commonly used states are considered as initial conditions for the harmonic oscillator: the Fock state, the thermal state, and the coherent state, each of which may be applicable in different situations. As might be expected, the behavior of the two-level system changes dramatically depending on the initial oscillator state. We have verified the behavior obtained in the adiabatic approximation against a numerical solution of the full Hamiltonian. Provided the ratio $\omega_0/\Omega$ is made large enough, the agreement is excellent over the short time scales which are likely to be experimentally accessible; the value $\omega_0/\Omega=10$ gives quantitative agreement over several dozen periods of oscillation. Time evolution in the adiabatic approximation shows a rich variety of behavior which we demonstrate and classify. 

Throughout this section the quantity to be examined is the probability of obtaining the state $\lvert - \rangle$ as a function of time, $P(-,t)$. The initial state of the two-level system is taken to be $\lvert - \rangle$, and the initial state of the oscillator is given in the displaced basis corresponding to the state $\lvert - \rangle$. This situation might be obtained, for example, by turning on the bias voltage $V_g$ between the NR and the CPB, tuning the CPB gate bias voltage $V_b$ so that the net bias is away from the degeneracy point, and allowing the CPB to relax to its ground state. Preparation of the oscillator state would vary depending on the type of state desired (Fock, thermal, or coherent); the discussion of each state includes some indication of how that state might be prepared. At time $t=0$ the microwave field is switched on and the system begins to evolve in time. Other initialization schemes might be imagined, but starting the oscillator in the displaced basis simplifies the mathematics.

An important point to keep in mind when comparing the results presented here with results from the JCM is that they are measured in different bases relative to their respective Hamiltonians. In our notation, the state $\lvert - \rangle$ is an eigenstate of the two-level-system operator $\hat{\sigma}_z$ which is associated with the interaction term of the Hamiltonian. In the JCM the initial state and the measured state of the atom are typically chosen to be eigenstates of the bare atomic Hamiltonian rather than the interaction Hamiltonian. These are natural choices for the corresponding physical systems, but it should be kept in mind that a degree of caution must be used in comparing the two models.

\begin{figure}
\includegraphics[scale=1]{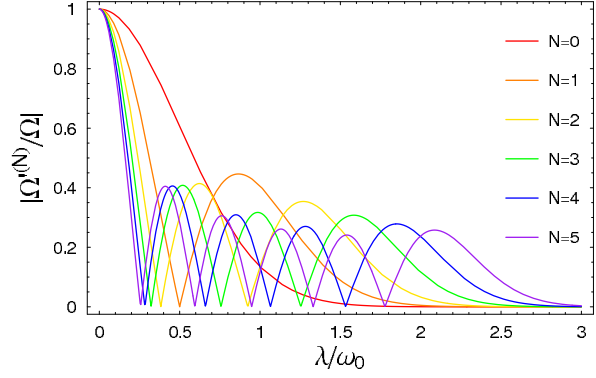}
\caption{\label{fig:freq_renorm}(Color online) Magnitude of the renormalized frequencies given by Eq.\ \eqref{freq1} for values of $N=0,1,\ldots,5$.}
\end{figure}

The simplest dynamical behavior is obtained when the oscillator begins in a displaced Fock state, such that $\lvert \psi(0) \rangle = \lvert - \rangle \otimes \lvert N_- \rangle$. The time evolution of the probability to find the two-level system in the initial state $\lvert - \rangle$ is given by
\begin{equation}\label{fock}
\begin{split}
P(-, t) &= \lvert \langle -, N_- \vert \psi(t) \rangle \rvert^2 \\
&= \cos^2(\Omega^{\prime (N)}t/2) ,
\end{split}
\end{equation}
where 
\begin{equation}\label{freq1}
\Omega^{\prime (N)} = \Omega \langle N_- \vert N_+ \rangle = \Omega e^{-2 \lambda^2 / \omega_0^2} L_N [(2 \lambda / \omega_0)^2] .
\end{equation}
The CPB undergoes Rabi oscillations (due to the external driving field), with a frequency which is modified by the strength of the coupling to the resonator; following Ref.\ \onlinecite{Leggett:1987} we refer to this as ``adiabatic renormalization.'' \footnote{It may be interesting to note that $\Omega^{\prime (0)}$ may be obtained from Eq.\ (3.20) of Ref.\ \onlinecite{Leggett:1987} by setting the spectral density $J(\omega)=\delta(\omega-\omega_0)$: that is, by treating the NR as a single-mode bath causing decoherence of the CPB.} Fig.\ \ref{fig:freq_renorm} shows a plot of the renormalized frequencies \eqref{freq1} versus the coupling strength $\lambda/\omega_0$ for small values of $N$. Unlike the Rabi frequencies obtained in the JCM, the frequencies found here are not monotonic functions of the coupling strength $\lambda$ or of $N$ for $N>0$. 

One unexpected feature of these frequencies is the zeros which occur as the coupling strength increases. Comparison with a numerical solution of the full Hamiltonian shows that the adiabatic approximation breaks down to some extent around the ``critical points'' in the coupling strength at which the renormalized Rabi frequency goes to zero. This point is discussed in Ref.\ \onlinecite{Crisp:1992}, and a higher-order formula is derived which is valid even near the critical points. Some caution must therefore be used in making predictions for Fock state initial conditions from our lowest-order formula. However, Fock states are highly non-classical states and although some methods for preparing such states in NRs have been proposed, \cite{Irish:2003,Santamore:2004} the experiments appear difficult. This paper is primarily concerned with time evolution from more realistic initial states which involve some distribution of number states. The distribution reduces the contribution from any given number state, and our numerical studies indicate that the approximation works well even when states are included which have critical points near a given coupling strength.

The next initial condition for the oscillator which we consider is a thermally-occupied state, sometimes also referred to as a chaotic state, in the displaced basis. This type of state is expected for an oscillator in thermal equilibrium with its environment, such as the NRs studied in the recent experiments of LaHaye et al. \cite{LaHaye:2004} The thermal state is a fully mixed state which must be described by a density matrix rather than a state vector. The two-level system is again taken to be initialized in the state $\lvert - \rangle$, so that the initial density matrix for the coupled system is given by
\begin{equation}
\rho_{\text{th}}(0) = \lvert - \rangle \langle - \rvert \otimes \sum_N p_{\text{th}}(N)\lvert N_- \rangle \langle N_- \rvert ,
\end{equation}
where
\begin{equation}
p_{\text{th}}(N) = \frac{1}{(1 + \langle N \rangle)(1 + 1/\langle N \rangle)^N} 
\end{equation}
and $\langle N \rangle = (e^{\hbar \omega_0 / k_B T} - 1)^{-1}$ is the average number of quanta in the oscillator at the temperature $T$.

Assuming that the system is weakly coupled to the thermal environment so that the influence of the environment is restricted to determining $\langle N \rangle$ (i.e., decoherence is not included), the time evolution of the system is given by 
\begin{equation}
\begin{split}
\rho_{\text{th}}(t) &= e^{-iHt/\hbar} \rho_{\text{th}}(0) e^{iHt/\hbar} \nonumber\\
&= \sum_N p_{\text{th}}(N)[\cos^2(\Omega^{\prime (N)}t/2) \; \lvert -, N_- \rangle \langle -,  N_- \rvert + \sin^2(\Omega^{\prime (N)}t/2) \; \lvert +, N_+ \rangle \langle +, N_+ \rvert \\
&\qquad \qquad \qquad + i \sin(\Omega^{\prime (N)}t/2) \; \cos(\Omega^{\prime (N)}t/2) \; (\lvert -, N_- \rangle \langle +, N_+ \rvert - \lvert +, N_+ \rangle \langle -, N_- \rvert )] .
\end{split}
\end{equation}
The reduced density matrix for the CPB is obtained by tracing over the oscillator states:
\begin{equation}
\begin{split}
\rho^{\text{CPB}}_{\text{th}}(t) &= \sum_N p_{\text{th}}(N)[\cos^2(\Omega^{\prime (N)}t/2) \; \lvert - \rangle \langle - \rvert + \sin^2(\Omega^{\prime (N)}t/2) \; \lvert + \rangle \langle + \rvert \\
&\qquad \qquad \qquad + i \sin(\Omega^{\prime (N)}t/2) \; \cos(\Omega^{\prime (N)}t/2) \; \langle N_- \vert N_+ \rangle \; (\lvert - \rangle \langle + \rvert - \lvert + \rangle \langle - \rvert ) ] .
\end{split}
\end{equation}
From the reduced density matrix the probability of obtaining the state $\lvert - \rangle$ is found to be
\begin{equation}\label{thermal}
\begin{split}
P_{\text{th}}(-, t) &= \langle - \vert \rho^{\text{CPB}}_{\text{th}}(t) \vert - \rangle \\
&= \sum_N p_{\text{th}}(N) \cos^2(\Omega^{\prime (N)}t/2) .
\end{split}
\end{equation}

If no further approximations are made, the sum in Eq.\ \eqref{thermal} requires numerical evaluation. The parameter space is complicated, but at least three qualitatively distinct regimes of behavior may be found, characterized by the oscillator temperature and the coupling strength. In the very low temperature regime, $\langle N \rangle \approx 0.01$, the behavior consists of ordinary Rabi oscillations with a frequency renormalized by the coupling to the ground state of the oscillator. This renormalization becomes significant for relatively large coupling strengths, $\lambda/\omega_0 \approx$ 0.1--1.

\begin{figure*}
\includegraphics[scale=1]{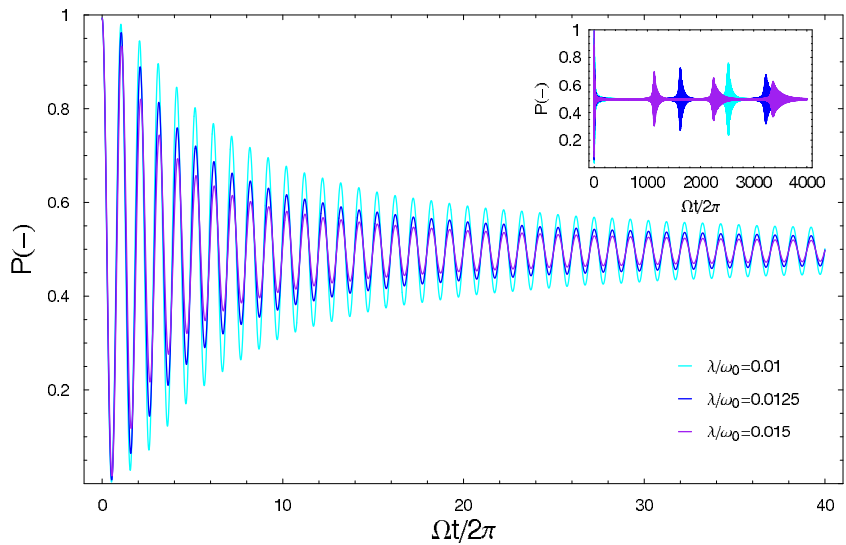}
\caption{\label{fig:hightemp_coupling_dep}(Color online) Behavior of the two-level-system occupation probability in the high-temperature ($\langle N \rangle = 100$), weak-coupling regime for short times (main figure) and long times (inset). Throughout the paper all time-dependent plots have $\Omega/\omega_0=1/10$.}
\end{figure*}

However, due to the large width in $N$ of the thermal distribution, the different frequencies $\Omega^{\prime (N)}$ involved in the time series \eqref{thermal} tend to interfere with each other, producing a decay in the amplitude of the oscillations. This effect, known as a ``collapse'' in quantum optics, \cite{Cummings:1965, Eberly:1980} dominates the short-time behavior at higher temperatures as long as $\lambda/\omega_0$ is not too large. At longer times, the discrete nature of the spectrum allows a partial rephasing of the oscillations, resulting in ``revivals'' of the oscillation amplitude. \cite{Eberly:1980} These phenomena are illustrated in Fig.\ \ref{fig:hightemp_coupling_dep}. Several features are worth noting. First is the nonzero amplitude in the collapse region; this residual amplitude decreases with increasing temperature. Second, the collapse time and the times at which the revivals occur depend on the coupling strength: both the collapse time and the revival time shorten as the coupling strength increases. Finally, the amplitude of the revivals decreases and their width increases as time goes on and the rephasing becomes less complete. Some of these features can be understood from a simple analytical approximation introduced in the next section.

\begin{figure}
\includegraphics[scale=1]{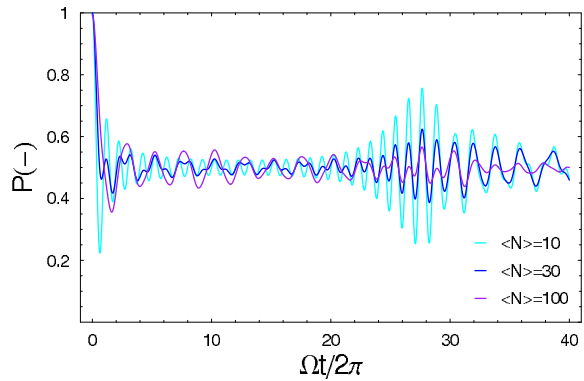}
\caption{\label{fig:revival_temp_dep}(Color online) Behavior of the first revival region as a function of temperature, with $\lambda/\omega_0 = 0.1$.}
\end{figure}

As the coupling strength is increased at a given temperature the behavior of the two-level system becomes increasingly erratic. Shorter revival times cause successive revivals to overlap and interfere so that the time evolution appears irregular. The coupling strength at which the irregularity emerges is closely tied to the temperature: the higher the temperature, the smaller the coupling strength needed to produce irregular behavior. Higher temperature also results in decreased revival amplitude: larger $\langle N \rangle$ corresponds to a larger number of frequencies in the sum, which in turn causes the rephasing to be less complete. However, a signature of the revivals persists in the form of a return to the bare Rabi frequency even at temperatures high enough that the behavior appears random and the revival amplitude is essentially washed out. Figure \ref{fig:revival_temp_dep} illustrates the lapse into erratic behavior and the persistent revival signature.

The variation of behavior in the thermal-state adiabatic approximation, from frequency renormalization to distinct collapse and revival dynamics to apparent randomness, contrasts with the findings of Knight and Radmore on the same type of system within the RWA. \cite{Knight:1982} Although they distinguish collapse and revival regions, the behavior within those regions appears erratic, reminiscent of that found above for large coupling strengths. Clear and well-defined revival pulses do not occur in the JCM for any parameter values if the oscillator begins in a thermal state. In fact, the basic shape of the time-evolution curve is invariant in the JCM, although the time scale and amplitude may change. The distinct revival areas found in the adiabatic approximation for smaller coupling strengths more closely resemble those obtained in the coherent-state JCM. \cite{Eberly:1980,Narozhny:1981} 

Finally we turn to the case in which the oscillator begins in a displaced coherent state. Coherent states are considered to be the quantum states of the harmonic oscillator which most closely approach the classical limit; it is expected that a driven NR is approximated by a coherent state when the external driving dominates thermal fluctuations. The initial condition for the coupled system may be written as the pure state
\begin{equation}
\lvert \psi_{\text{coh}}(0) \rangle = \lvert - \rangle \otimes e^{-\lvert \alpha \rvert^2/2} \sum_{N=0}^{\infty} \frac{\alpha^N}{(N!)^{1/2}} \lvert N_- \rangle ,
\end{equation}
where $\alpha$ is the (complex) amplitude of the coherent state, and we will define $\langle N \rangle = \lvert \alpha \rvert^2$.

The time evolution of the corresponding density matrix $\rho_{\text{coh}}(0)=\lvert \psi_{\text{coh}}(0)\rangle \langle \psi_{\text{coh}}(0) \rvert$ may be calculated in the same way as before, with the resulting time-dependent probability
\begin{equation}\label{coherent}
P_{\text{coh}}(-,t) = \sum_{N=0}^{\infty} p_{\text{coh}}(N) \cos^2 (\Omega^{\prime(N)} t/2) ,
\end{equation}
where $p_{\text{coh}}(N) = e^{-\langle N \rangle} \langle N \rangle^N/N!$. Equation~\eqref{coherent} has the same form as Eq.\ \eqref{thermal} with the weighting function for the thermal distribution replaced by the weighting function for the diagonal elements of the coherent state.

As before, the sum in Eq.\ \eqref{coherent} requires numerical evaluation if no further approximations are made. However, the qualitative behavior is more difficult to classify. The regime $\langle N \rangle \approx 0.01$ behaves much like the same regime in the thermal case. Frequency renormalization is the dominant effect, visible for relatively large values of $\lambda/\omega_0$.

\begin{figure*}
\includegraphics[scale=1]{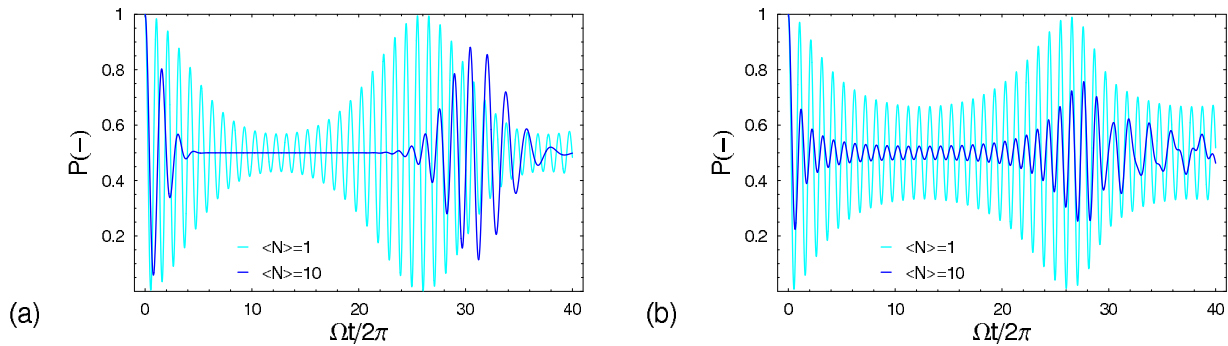}\caption{\label{fig:coh_vs_th}(Color online) Collapse and first revival for (a) coherent and (b) thermal states in a regime of regular behavior, $\lambda/\omega_0 = 0.1$. For large $\langle N \rangle$, the coherent-state model results in a complete collapse, with no residual oscillation amplitude in the collapse region; this is not true for the thermal-state model.}
\end{figure*}

For values of $\langle N \rangle \gtrsim 10$ and fairly small coupling strengths $\lambda$, collapses and revivals appear which look similar to those found in the JCM. \cite{Eberly:1980} The collapses are complete, with virtually no residual amplitude in the collapse region, unlike the above results for the thermal state. For small values of $\langle N \rangle$ the coherent state does not have a wide enough spread in frequencies to create a complete collapse, resulting in residual oscillations in the collapse region. Figure \ref{fig:coh_vs_th} compares the coherent-state behavior with the thermal-state behavior. As in the thermal case, a simple analytic approximation that will be presented in the next section explains some of these features.

\begin{figure*}
\includegraphics[scale=1]{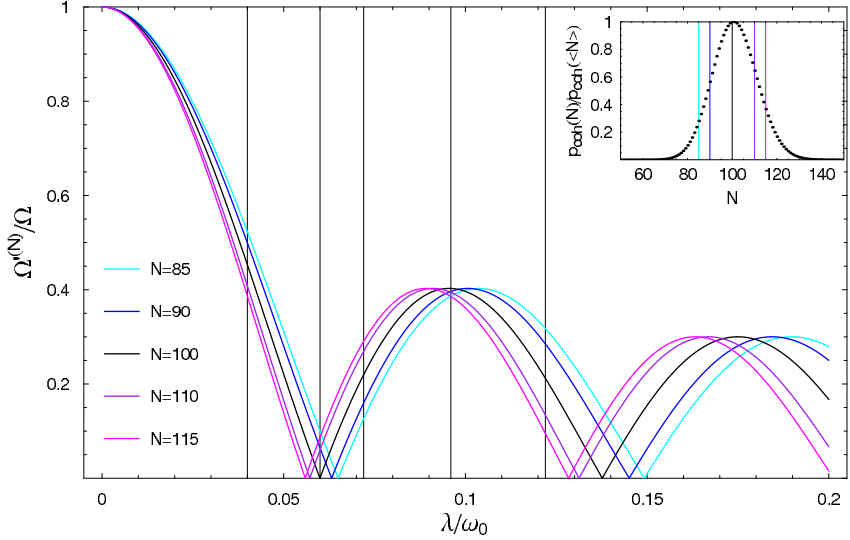}
\caption{\label{fig:hightemp_freq_func}(Color online) Inset: Normalized coherent-state probability distribution as a function of $N$ with $\langle N \rangle=100$. Vertical lines correspond to the values of $N$ whose frequencies are plotted in the main figure. Main figure: Frequency renormalization as a function of coupling strength for some representative values of $N$. Vertical lines correspond to the coupling strengths used in Fig.\ \ref{fig:coh_hightemp_weird}. }
\end{figure*}

\begin{figure*}
\includegraphics[scale=1]{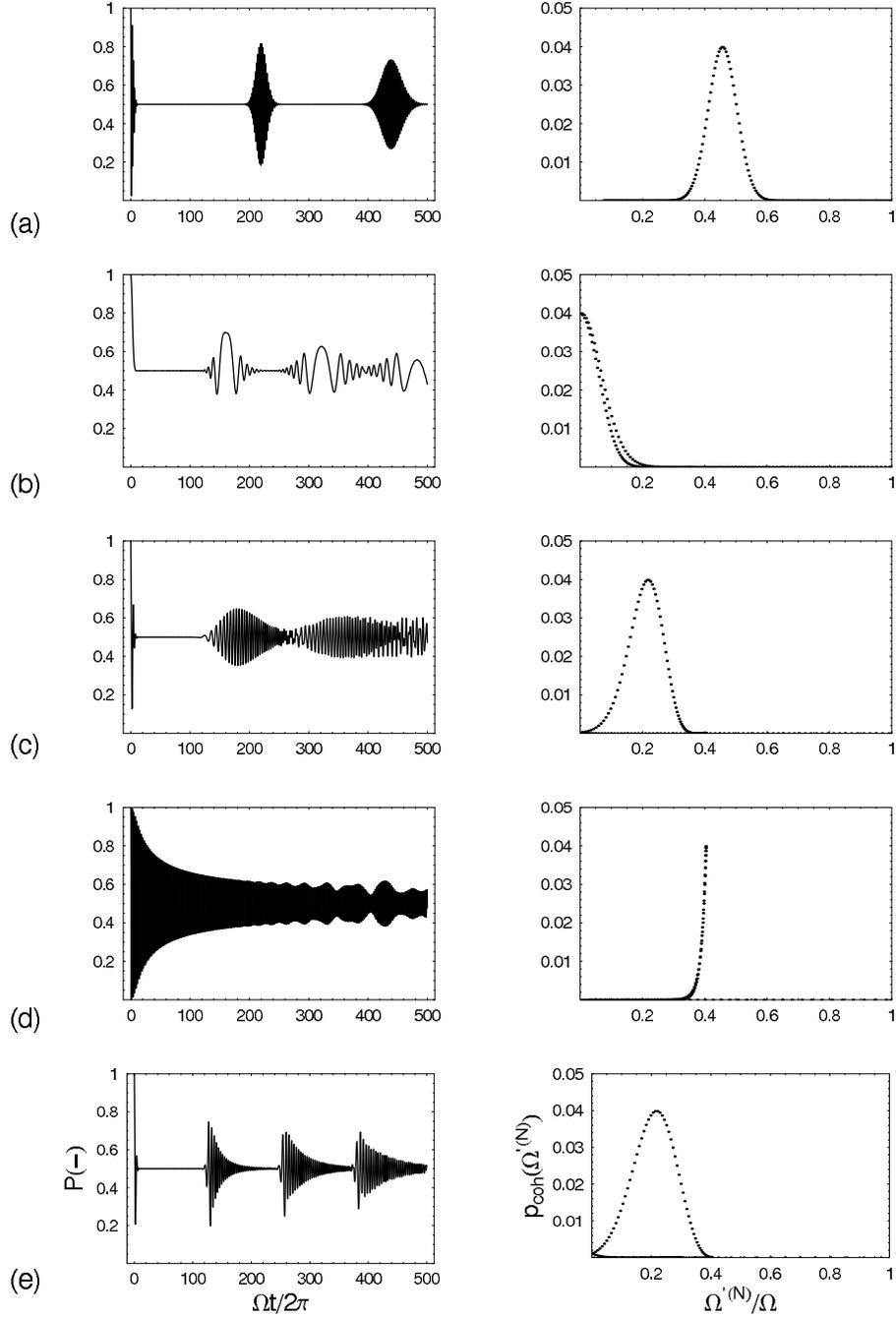}\caption{\label{fig:coh_hightemp_weird} A few samples of the unusual behavior which appears in the coherent-state case with $\langle N \rangle=100$ (left-hand side) and the associated weighted frequency distributions $p_{\text{coh}}(\Omega^{\prime (N)})$ (right-hand side). The coupling strengths used are $\lambda/\omega_0=$ (a) $0.04$, (b) $0.06$, (c) $0.072$, (d) $0.096$, (e) $0.122$.}
\end{figure*}

At larger coupling strengths, however, the coherent-state behavior does not necessarily lapse into irregularity as in the thermal case. For large $\langle N \rangle$, some unexpected results occur as the coupling is increased. The explanation for this lies in the non-monotonic dependence of the modified Rabi frequencies $\Omega^{\prime (N)}$ on both $\lambda$ and $N$. $\Omega^{\prime (N)}$ is plotted as a function of $\lambda/\omega_0$ in Fig.\ \ref{fig:hightemp_freq_func}; the values of $N$ illustrated are chosen from the coherent-state distribution with $\langle N \rangle = 100$ (inset) in order to give a sense of how the spread in frequency corresponds to the distribution in $N$. Figure \ref{fig:coh_hightemp_weird} illustrates the resulting behavior of the two-level system in both time and frequency space. The left-hand side shows plots of $P(-,t)$ with $\langle N \rangle=100$ for several values of $\lambda/\omega_0$. Vertical lines in Fig.\ \ref{fig:hightemp_freq_func} correspond to those values of $\lambda/\omega_0$ for which time series are given in Fig.\ \ref{fig:coh_hightemp_weird}. Considering the coherent-state weighting function $p_{\text{coh}}(N)$ as a function of frequency yields a Fourier-transform-like distribution $p_{\text{coh}}(\Omega^{\prime (N)})$, which gives the amplitude of each frequency in the sum $P(-,t)$. The weighted frequency distribution corresponding to each time series is plotted on the right-hand side of Fig.\ \ref{fig:coh_hightemp_weird}. 

At zero coupling strength the weighted frequency distribution consists of a delta function located at $\Omega^{\prime (N)}/\Omega=1$. As the coupling strength is increased, the center of the distribution shifts toward smaller frequencies and the shape spreads out and becomes approximately Gaussian, which results in well-defined collapse and revival regions [Fig.\ \ref{fig:coh_hightemp_weird}(a)]. When the coupling strength approaches the critical point for the center of the distribution ($\Omega^{\prime (\langle N \rangle)} = 0$), the function $p_{\text{coh}}(\Omega^{\prime (N)})$ begins to fold back on itself, resulting in a very fast collapse and strangely shaped revivals [Fig.\ \ref{fig:coh_hightemp_weird}(b)]. The center of the distribution function then shifts back to higher frequencies as the second ``hump'' of the renormalized frequency function $\Omega^{\prime (N)}$ is traversed (Fig.\ \ref{fig:hightemp_freq_func}), and the shape again becomes almost Gaussian. Correspondingly the time-dependent probability appears more regular [Fig.\  \ref{fig:coh_hightemp_weird}(c)]. At the peak of the second hump of $\Omega^{\prime (N)}$ ($\lambda/\omega_0 \approx 0.095$ in Fig.\ \ref{fig:hightemp_freq_func}) there is very little dispersion in the frequencies corresponding to different values of $N$, resulting in a nearly delta-function distribution for $p_{\text{coh}}(\Omega^{\prime (N)})$ and a very slow collapse [Fig.\ \ref{fig:coh_hightemp_weird}(d)]. The frequency distribution function then ``bounces'' back toward low frequencies. Deviations from a Gaussian distribution show up in the time-dependent probabilities as altered revival pulse shapes [Fig.\ \ref{fig:coh_hightemp_weird}(e)]. The behavior evolves between the examples shown here in a continuous manner, although in places it varies quite rapidly with $\lambda/\omega_0$.

The previous discussion highlights the complexity of the coherent-state adiabatic model. It should be noted that some of the phenomena seen above are blurred out at smaller values of $\langle N \rangle$. This may be understood by noting that two adjacent curves $\Omega^{\prime (N)}$ and $\Omega^{\prime (N+1)}$ diverge more strongly at small values of $N$: compare Fig.\ \ref{fig:freq_renorm}, with $N \approx 1$, and Fig.\ \ref{fig:hightemp_freq_func}, with $N \approx 100$. Thus the weighted frequency distributions $p_{\text{coh}}(\Omega^{\prime (N)})$ have a wider spread for small values of $\langle N \rangle$ despite the fact that the coherent-state number distribution is narrower in $N$ for small $\langle N \rangle$. The wider spread in frequency space for smaller $\langle N \rangle$ results in more erratic behavior, without the returns to regularity demonstrated in Fig.\ \ref{fig:coh_hightemp_weird}.

A similar analysis in frequency space also explains why the thermal-state dynamics appears less complex than the coherent-state dynamics. As a function of $N$, the maximum of the thermal state distribution is fixed at $N=0$ regardless of the value of $\langle N \rangle$, while the maximum of the coherent state distribution is given by $\langle N \rangle$. The frequency shift in the state $N=0$ is monotonic as a function of $\lambda/\omega_0$ and weak compared to the shift for higher values of $N$ (Fig.\ \ref{fig:freq_renorm}). Thus the maximum of the weighted frequency distribution $p_{\text{th}}(\Omega^{\prime (N)})$ for the thermal state shifts less dramatically than the maximum of $p_{\text{coh}}(\Omega^{\prime (N)})$ for the coherent state, and the changes in the shape of the distribution are less pronounced. 

Applying the frequency distribution function analysis to the JCM shows how the complicated dependence on $N$ and $\lambda$ of the frequencies $\Omega^{\prime (N)}$ in the adiabatic approximation leads to a much richer variety of behavior than that found in the JCM. The Rabi frequencies in the JCM change monotonically with $N$ and $\lambda$ in such a way that the frequency distribution function changes in width but never in shape. Changes in $\langle N \rangle$ and $\lambda$ in the JCM result in changes in the amplitude and time scale of the evolution; however, the shape of the time series is unaltered. The large variation of behavior with $\langle N \rangle$ and $\lambda$ is a distinctive feature of the adiabatic approximation, not found in the JCM.

\section{The weak coupling limit}\label{Weak coupling}

Although the adiabatic approximation allows evaluation of the time-dependent behavior of the two-level system, the results for thermal or coherent oscillator states are given by infinite sums which must be numerically evaluated. This is also true in the JCM, although integral approximation techniques have been used to derive some approximate analytic expressions. \cite{Eberly:1980,Narozhny:1981} An interesting feature of the adiabatic approximation is that taking the weak coupling limit allows closed-form evaluation of the time evolution for both thermal and coherent initial states. 

\begin{figure}
\includegraphics[scale=1]{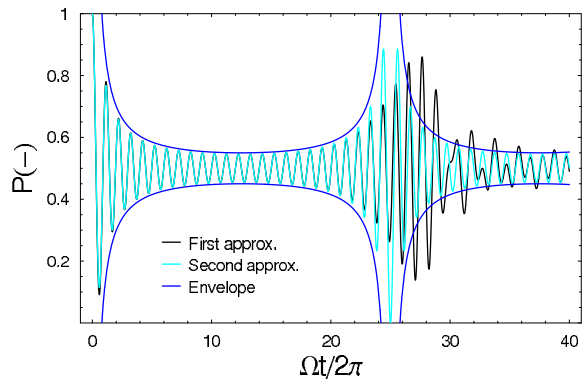}
\caption{\label{fig:th_sec_approx}(Color online) Comparison of the first (adiabatic) approximation [Eq.\ \eqref{thermal}] and the second (weak-coupling) approximation [Eq.\ \eqref{thapprox2}] for the thermal-state model. Also shown is the envelope function derived from Eq.\ \eqref{thapprox3} by neglecting the rapidly-oscillating factor $\sin(\Omega t)$. Parameters used are $\langle N \rangle=5$ and $\lambda/\omega_0=0.1$. Although these values are near the limit of validity of the approximations used, the agreement up to the first revival is excellent.}
\end{figure}

For small values of $(\lambda/\omega_0)^2$ the modified Rabi frequencies given in Eq.\ \eqref{freq1} may be approximated as \cite{Abramowitz}
\begin{equation}\label{laguerreapprox}
\Omega^{\prime (N)} \approx \Omega[1-(N+1/2)(2 \lambda / \omega_0)^2] .
\end{equation} 
This approximation enables the sums in Eqs.\ \eqref{thermal} and \eqref{coherent} to be carried out. \cite{Abramowitz} For the thermal state, Eq.\ \eqref{thermal} becomes
\begin{equation}\label{thapprox2}
P_{\text{th}}(-, t) \approx \frac{1}{2} + \frac{1}{2} \frac{\cos (\Omega t) \cos [\Omega(2 \lambda/\omega_0)^2 t/2] + (1 + 2 \langle N \rangle) \sin (\Omega t) \sin [\Omega(2 \lambda/\omega_0)^2 t/2]}{1 + 4 \langle N \rangle (1 + \langle N \rangle) \sin^2 [\Omega(2 \lambda / \omega_0)^2 t/2]} .
\end{equation}
A comparison of Eqs.\ \eqref{thermal} and \eqref{thapprox2} is shown in Fig.\ \ref{fig:th_sec_approx}. It is apparent that even for relatively large values of $\lambda/\omega_0$ the approximation to the sum works quite well in the initial collapse region and captures the general qualitative behavior of the function. In the limit $\langle N \rangle \gg 1$, Eq.\ \eqref{thapprox2} reduces to
\begin{equation}\label{thapprox3}
P_{\text{th}}(-,t) \approx \frac{1}{2} + \frac{\sin \Omega t}{4 \langle N \rangle \sin[\Omega(2 \lambda/\omega_0)^2 t/2]} .
\end{equation}
From this formula some of the characteristics discussed in the previous section are immediately evident. Revivals occur when $\Omega (2 \lambda/\omega_0)^2 t/2 = \pi m$ for $m=1,2,\ldots$, giving a revival time $\tau_r = 2 \pi/[\Omega (2 \lambda/\omega_0)^2]$ which decreases with increasing $\lambda/\omega_0$ as expected. The minimum oscillation amplitude in the collapse region scales as $1/\langle N \rangle$. However, Eq.\ \eqref{thapprox3} diverges as $t \to 0$ as well as at the revival times, so it is not useful in predicting the shape of the collapse envelope at short times or the nature of the revival envelope function.

\begin{figure}
\includegraphics[scale=1]{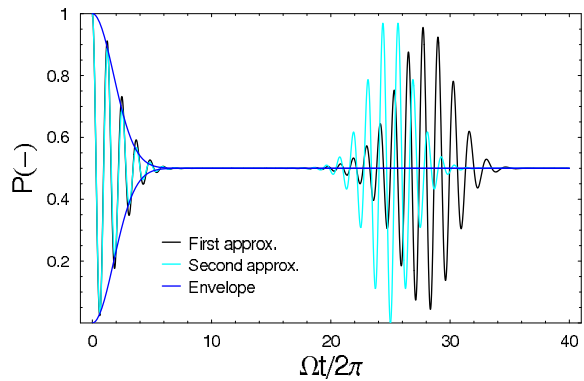}
\caption{\label{fig:coh_sec_approx}(Color online) As for Fig.\ \ref{fig:th_sec_approx}, but for the coherent-state model [Eqs.\ \eqref{coherent} and \eqref{cohapprox2}]; all parameters are identical. The envelope shown is the Gaussian collapse function given by Eq.\ \eqref{cohapprox3} without the rapidly-oscillating cosine factor.}
\end{figure}

Within the weak-coupling approximation, the coherent-state evolution may be evaluated as well. Equation \eqref{coherent} yields the sum
\begin{equation}\label{cohapprox2}
P_{\text{coh}}(-, t) \approx \tfrac{1}{2} + \tfrac{1}{2} e^{-2 \langle N \rangle \sin^2[\Omega (2 \lambda/\omega_0)^2 t/2]} \cos \lbrace \langle N \rangle \sin[\Omega (2 \lambda/\omega_0)^2 t] - [1-(2 \lambda/\omega_0)^2/2] \Omega t \rbrace ,
\end{equation}
which in the limit $\Omega (2 \lambda/\omega_0)^2 t \ll 1$ reduces to
\begin{equation}\label{cohapprox3}
P_{\text{coh}}(-,t) \approx \tfrac{1}{2} + \tfrac{1}{2} e^{-[\langle N \rangle \Omega^2 (2 \lambda/\omega_0)^4/2]t^2} \cos\lbrace[(\langle N \rangle + \tfrac{1}{2})(2 \lambda/\omega_0)^2 - 1] \Omega t\rbrace .
\end{equation}
The last form demonstrates the short-time Gaussian collapse envelope which is also found in the coherent-state JCM. \cite{Cummings:1965,Eberly:1980,Narozhny:1981} However, the full expression given in Eq.\ \eqref{cohapprox2} is necessary in order to obtain revivals. Notice that the revival time $\tau_r$ is the same as in the thermal case. A comparison of Eqs.\ \eqref{coherent} and \eqref{cohapprox2} is shown in Fig.\ \ref{fig:coh_sec_approx}. As in the thermal state case, the approximation to the sum works well in the initial collapse region.

This approximation highlights once again the role of the functional form of the modified Rabi frequencies \eqref{freq1} in controlling the time evolution. Rabi frequencies linear in $N$ are obtained in a model similar to the JCM but involving three atomic levels and two photons. \cite{Knight:1986} Since all the frequencies involved are integer multiples of the bare Rabi frequency, the interfering oscillations rephase completely, leading to perfectly periodic, full-amplitude revivals for both coherent-state and thermal-state initial conditions. As demonstrated above, the renormalized Rabi frequencies in the adiabatic approximation are linear in $N$ to first order. This is not the case in the usual two-level JCM, which yields Rabi frequencies which go as the square root of $N$. The closer approach to linearity in $N$ explains why clear revivals may be found in the adiabatic model even for a thermal state, while the thermal-state JCM always produces erratic behavior.

\section{Experimental prospects}\label{Experimental}

The primary requirement for experimental exploration of physics in the adiabatic regime is strong coupling at a large detuning between the two-level system and the harmonic oscillator. Atom-cavity systems typically have coupling strengths $\lambda/\omega_0 \approx 10^{-7}$ -- $10^{-6}$ at detunings of $\Delta/\omega_0 \approx 10^{-7}$ -- $10^{-5}$, and are well described by the RWA. \cite{Raimond:2001,Hood:2000} The adiabatic regime requires numbers several orders of magnitude larger, unlikely to be achieved with atoms. However, recent progress in solid-state systems suggests that experimental implementation of the model discussed here may be possible fairly soon. A system consisting of a CPB coupled to a superconducting transmission line has achieved the ``strong-coupling limit'' of CQED, in which coherent dynamics occurs faster than the decoherence rates, as confirmed by spectroscopic \cite{Wallraff:2004} and time-domain \cite{Wallraff:2005} experiments. The coupling strength obtained at zero detuning was $\lambda/\omega_0 \approx 10^{-3}$, which is a significant improvement over atomic systems. However, since the coupling is through the electric dipole moment it may be difficult to increase the coupling strength by the two orders of magnitude needed to get out of the RWA regime, and significant coupling at large detunings is unlikely. 

Coupling strengths much larger than those possible with dipole coupling may be achieved with capacitive or inductive couplings. Some recent experimental results on a flux qubit inductively coupled to a superconducting quantum interference device (SQUID) that acts as both a measuring device and a quantum harmonic oscillator appear very promising. \cite{Chiorescu:2004} Coherent oscillations in the qubit were observed with Rabi frequencies ranging from very small to as large as the qubit splitting frequency. Since the oscillator frequency was about half the qubit splitting frequency, an analysis similar to that given below for the CPB-NR system yields values of $\Omega/\omega_0 \lesssim 2$. Given the coupling strength of $\lambda/\omega_0 \approx 0.1$, the adiabatic regime is already within the reach of this system. 

Although it has not yet been experimentally demonstrated, the system consisting of a CPB capacitively coupled to a NR appears to be another potential candidate for achieving the adiabatic regime. The remainder of this section is devoted to an analysis of the circumstances under which this would be possible. 

The Hamiltonian is given by \cite{Irish:2003}
\begin{align}\label{Hsystem}
H_{\text{TOTAL}}&=H_{\text{CPB}}+H_{\text{NR}}+H_{\text{int}}\\
H_{\text{CPB}}&=4E_C(n_g-\tfrac{1}{2})\hat{\sigma}_z-\tfrac{1}{2}E_J\hat{\sigma}_x\\
H_{\text{NR}}&=\hbar\omega_0 \hat{a}^\dagger \hat{a} \\
H_{\text{int}}&=\hbar \lambda (\hat{a}^\dagger + \hat{a})\hat{\sigma}_z ,
\end{align}
where $\hat{a}^\dagger,\hat{a}$ are harmonic-oscillator raising and lowering operators which act on the NR; $\hat{\sigma}_z, \hat{\sigma}_x$ are Pauli spin matrices operating in the charge basis of the CPB; $n_g=(C_b V_b + C_g V_g)/2e$ where $C_b$ and $V_b$ are the CPB biasing capacitance and voltage and $C_g$ and $V_g$ are the capacitance and voltage between the NR and the CPB; $E_C$ and $E_J$ are the Coulomb and Josephson energies; $\omega_0$ is the NR oscillator frequency in the absence of coupling; and  $\lambda=-4E_C n_g^{\text{NR}} \Delta x_{\text{ZP}}/\hbar d$ where $n_g^{\text{NR}} = C_g V_g /2e$, $\Delta x_{\text{ZP}}=\sqrt{\hbar/(2m\omega_0)}$ is the zero-point position uncertainty of the NR with effective mass $m$, and $d$ is the distance between the NR and the CPB.

One way of reducing this Hamiltonian to the form of Eq.\ \eqref{Hamiltonian} is to bias the Cooper-pair box to the degeneracy point, where $n_g = 1/2$. However, typical values of $E_J/h$ are on the order of several gigahertz, whereas the highest reported nanomechanical resonator frequency is about $1$ GHz. \cite{Huang:2003} The approximation we have derived in this paper is based on the assumption that the effective splitting frequency of the two-level system is much smaller than the frequency of the oscillator, $\Omega \ll \omega_0$, which would be difficult to satisfy for degeneracy-point biasing. 

However, an effective Hamiltonian of the correct form which satisfies the condition $\Omega \ll \omega_0$ may be found given a few reasonable assumptions. Physically, this involves biasing the CPB well away from degeneracy so that $4E_C(n_g-1/2) \gg E_J/2$ and applying an oscillating voltage to the CPB bias gate, a procedure which is used in performing spectroscopy. \cite{Lehnert:2003} Assuming that the frequency of the oscillating voltage is equal to the splitting frequency of the box and that the amplitude of the oscillating voltage is small, the Hamiltonian may be approximated by Eq.\ \eqref{Hamiltonian} with $\Omega = 8g E_J E_C(1/2-n_g)/\lbrace[8 E_C(1/2-n_g)]^2 + E_J^2 \rbrace$, where $g=4E_C C_b V_b^{\text{ac}}/2e\hbar$ is the Rabi frequency induced by the oscillating voltage. The Appendix contains a detailed derivation of this approximation.

Achievable parameters for an experiment of this type align well with the regime in which the adiabatic approximation is valid. Starting with a NR frequency \cite{Huang:2003} $\omega_0/2\pi = 1~\text{GHz}$, taking $\Omega/\omega_0=1/10$ requires an effective CPB Rabi frequency $\Omega/2\pi = 100~\text{MHz}$. With CPB parameters $E_C/h = 20~\text{GHz}$, $E_J/h = 7~\text{GHz}$, and $n_g=1/4$, the Rabi frequency induced by the oscillating voltage should be about $0.5~\text{GHz}$. Assuming a bias-gate capacitance of $C_b = 10~\text{aF}$, the required amplitude for the oscillating voltage is $V_b^{\text{ac}} = 0.2~\text{mV}$, which may be achieved easily. Note that these numbers also satisfy the requirements for derivation of the effective Hamiltonian which are given in the Appendix. Since the coupling is capacitive, $\lambda/\omega_0$ is limited by how small the distance between the NR and the CPB island can be made and by how large a voltage may be applied without damage to the NR. Values on the order of $\lambda/\omega_0 \approx 10^{-2}$ -- 1 should be possible. All of these parameters appear to be well within the reach of present technology.

Unfortunately the dephasing times for coherent oscillations in a CPB which have been measured so far are quite short. Vion \textit{et al.} \cite{Vion:2002} found a dephasing time of $0.5~\mu\text{s}$ at the degeneracy point. As the bias voltage is tuned away from the degeneracy point, the dephasing time drops rapidly, \cite{Duty:2004} which is attributed to low-frequency charge noise. In order to see the effects predicted here, coherence times of several effective Rabi periods, on the order of several tens of nanoseconds, would be necessary; this would require significant improvement over current experiments. However, charge noise is not believed to be intrinsic to these systems, and advances in materials and fabrication may reduce the problem.

\section{Conclusion}\label{Conclusion}

The adiabatic approximation we have discussed in this paper provides a rich and robust framework for exploring spin-oscillator physics outside the rotating-wave approximation. Although it is derived under the assumption that the two-level splitting frequency $\Omega$ is much smaller than the oscillator frequency $\omega_0$, it works well even when $\Omega = \omega_0$; indeed, it provides a more accurate description at large coupling strengths than the RWA. The energy levels obtained from this model exhibit nonmonotonic dependence on the oscillator occupation number $N$ and the coupling strength $\lambda$. This leads to complicated time-dependent behavior of the two-level system, which may exhibit frequency renormalization, collapse and revival of coherent oscillations, or apparent randomness. Such behavior is quite sensitive to the initial state of the oscillator and the coupling strength in some parameter regimes.

Although the adiabatic approximation is not a new result, solid-state experiments currently underway provide motivation for a more thorough exploration of its consequences. The pursuit of quantum computing has catalyzed the development of new types of devices which act as artificial atoms. Given the success of atom-cavity experiments in demonstrating various characteristics of quantum behavior, it is not surprising that solid-state analogs are being pursued. Such systems have the capability to reach regimes, inaccessible to traditional atom-cavity systems, in which the RWA is no longer valid. This paper has demonstrated some of the complexity which may be encountered at large detuning and strong coupling.  

In particular, we have chosen to focus on a charge-based two-level system coupled to a nanomechanical resonator. Observation of the two-level system may offer some insight into the quantum nature of the resonator, just as atoms provide a sensitive probe for the nonclassical nature of electromagnetic fields. At fairly high resonator temperatures, the shape of the collapse of the coherent oscillations in the CPB may provide some information about the distribution of NR states: a thermal state gives a different envelope function than a coherent state. For either distribution, the shift from collapse dynamics to frequency renormalization would be a clear indication of near-ground-state cooling of the resonator. Finally, the observation of revivals, which are a strictly nonclassical phenomenon, would give evidence for the quantum nature of a macroscopic mechanical object. Such experiments appear to be nearly within the reach of current technology.

\begin{acknowledgments}
Helpful discussions with Keith Miller, Alex Hutchinson, Carlos Sanchez, Akshay Naik, and Marc Manheimer are gratefully acknowledged. E. K. I. acknowledges support from the National Physical Sciences Consortium.
\end{acknowledgments}

\appendix*
\section{Derivation of the effective CPB Hamiltonian}

The Hamiltonian for the Cooper-pair box in the basis of charge states is  \cite{Makhlin:2001}
\begin{equation}\label{CPBHam}
H_{\text{CPB}} = -4E_C(\tfrac{1}{2}-n_g)\hat{\sigma}_z-\tfrac{1}{2}E_J\hat{\sigma}_x ,
\end{equation}
where $\hat{\sigma}_z, \hat{\sigma}_x$ are Pauli spin matrices operating in the charge basis of the CPB; $n_g=C_b V_b/2e$ where $C_b$ and $V_b$ are the CPB biasing capacitance and voltage; and $E_C$ and $E_J$ are the Coulomb and Josephson energies. Eq.\ \eqref{CPBHam} may also be written in terms of the mixing angle $\eta \equiv \tan^{-1}\lbrace E_J/[8E_C(1/2 - n_g)]\rbrace$:
\begin{equation}
H_{\text{CPB}} = -\tfrac{1}{2} \Delta E(\eta) (\cos \eta \hat{\sigma}_{z} + \sin \eta \hat{\sigma}_{x}) ,
\end{equation}
where $\Delta E(\eta) = \sqrt{[8E_C(1/2 - n_g)]^2 + E_J^2}$. Alternatively, the Hamiltonian may be written in the diagonal form
\begin{equation}
\tilde{H}_{\text{CPB}} = -\tfrac{1}{2} \Delta E(\eta) \hat{\rho_z}
\end{equation}
where the Pauli operators which operate in the eigenbasis of $H_{\text{CPB}}$ are defined as
\begin{align}
\hat{\rho}_{z} &\equiv \cos \eta \hat{\sigma}_{z} + \sin \eta \hat{\sigma}_{x} \\
\hat{\rho}_{x} &\equiv \cos \eta \hat{\sigma}_{x} - \sin \eta \hat{\sigma}_{z} \\
\hat{\rho}_{y} &\equiv \hat{\sigma}_{y} .
\end{align}

An oscillating voltage $V_{b}^{\text{ac}}$ may be applied to the bias gate of the CPB, resulting in an additional term in the Hamiltonian \cite{Lehnert:2003} $H_{\text{ac}} = \hbar g \cos(\omega t) \hat{\sigma}_z$ where $g=4E_C C_b V_b^{\text{ac}}/2e\hbar$. In the eigenbasis of $H_{\text{CPB}}$ this becomes
\begin{equation}
\tilde{H}_{\text{ac}} = \tfrac{1}{2}\hbar g (e^{i \omega t} + e^{-i \omega t})[\cos \eta \hat{\rho}_{z} - \sin \eta(\hat{\rho}_{+} + \hat{\rho}_{-})] ,
\end{equation}
where $\hat{\rho}_{\pm} = \tfrac{1}{2}(\hat{\rho}_x \pm i\hat{\rho}_{y})$ are raising (lowering) operators for the CPB.

We will assume that $\omega \approx \Delta E(\eta)/\hbar$ so that the rotating wave approximation may be used to derive a time-independent effective Hamiltonian for the CPB with oscillating bias voltage. The first step is to transform into a reference frame which rotates about the $\hat{\rho}_{z}$ axis at the frequency $\omega$. This may be accomplished by the transformation $\tilde{H}_{\text{rot}} = \hat{U}^{\dag} \tilde{H} \hat{U} - i \hbar \hat{U}^{\dag} d\hat{U}/dt$ with $\hat{U} = \exp(i \omega t \hat{\rho}_{z}/2)$. Noting that $\hat{U}^{\dag} \hat{\rho}_{\pm} \hat{U} = \exp(\mp i\omega t) \hat{\rho}_{\pm}$, the Hamiltonian $\tilde{H}_{\text{CPB}} + \tilde{H}_{\text{ac}}$ transforms to
\begin{equation}\label{Hrot}
\tilde{H}_{\text{rot}} = -\tfrac{1}{2} \hbar \Delta \hat{\rho}_{z} + \tfrac{1}{2}\hbar g (e^{i \omega t} + e^{-i \omega t}) \cos \eta \hat{\rho}_{z} - \tfrac{1}{2} \hbar g \sin \eta (\hat{\rho}_{+} + \hat{\rho}_{-} + e^{-2 i \omega t} \hat{\rho}_{+} + e^{2i \omega t} \hat{\rho}_{-})
\end{equation}
where $\Delta \equiv \Delta E(\eta)/\hbar - \omega$ is the detuning between the CPB splitting frequency and the frequency of the oscillating bias voltage. As long as we are interested in motion on the timescale of $1/g \gg 1/\omega$, the time-dependent terms in Eq.\ \eqref{Hrot} may be neglected. With this approximation, and transforming back to the charge basis, we obtain
\begin{equation}\label{Happrox}
H_{\text{CPB}} \approx -\tfrac{1}{2} \hbar \Delta(\cos \eta \hat{\sigma}_{z} + \sin \eta \hat{\sigma}_{x}) - \tfrac{1}{2} \hbar g \sin \eta(\cos \eta \hat{\sigma}_{x} - \sin \eta \hat{\sigma}_{z}) .
\end{equation}
If we assume the CPB to be biased far from degeneracy such that $\sin \eta \ll \cos \eta$ and take the detuning $\Delta = 0$, we find an effective Hamiltonian for the CPB \footnote{For the sake of simplicity, the effective Hamiltonian is left in the rotating frame. The effect on the expectation value of $\hat{\sigma}_z$, which is all we consider in this paper, is neglible within the assumptions we have already made.} 
\begin{equation}\label{Heff}
H_{\text{CPB}}^{\text{eff}} = -\tfrac{1}{2} \hbar \Omega \hat{\sigma}_{x} 
\end{equation}
where $\Omega \equiv g \sin \eta \cos \eta$. The Hamiltonians for the NR and the interaction between the CPB and NR remain as given in Eq.\ \eqref{Hsystem}. Combining all the terms yields a Hamiltonian for the coupled system of the form of Eq.\ \eqref{Hamiltonian}.

\end{document}